\documentstyle[11pt,newpasp,twoside,epsf]{article}
\def\edcomment#1{\iffalse\marginpar{\raggedright\sl#1\/}\else\relax\fi}
\reversemarginpar

\begin{document}
\title{Formation of Binary Pulsars in Globular Clusters}
 \author{Frederic A.\ Rasio}
\affil{Department of Physics and Astronomy, Northwestern University, 
Evanston, IL 60208, USA}

\begin{abstract}
Close to 70 radio pulsars have now been discovered in 23 globular
clusters, with a record 22 pulsars observed in 47~Tuc alone.
Accurate timing solutions, including positions in the cluster,
are known for many of the pulsars.
These recent observations provide a unique opportunity
to re-examine theoretically the formation and evolution of recycled pulsars
in globular clusters. This brief review focuses on dynamical
exchange interactions between neutron stars and primordial binaries,
through which neutron stars can acquire intermediate-mass binary
companions. The later evolution of these intermediate-mass binaries
leads naturally to the two main types of binary millisecond pulsars
observed in clusters.
\end{abstract}

\section{Introduction}

The properties of globular cluster pulsars (see articles by Lorimer et al.\
and Ransom et al.\ in this volume) are rather surprising. While
some pulsars are single, the majority are in short-period binaries. Most of
the binaries have properties similar to those of the rare
``eclipsing binary pulsars'' seen in the Galactic disk population 
(see Nice et al.\ 2000 for a review).
These systems have extremely short orbital periods, $P_b\sim1-10\,$hr,
circular orbits, and
very low-mass companions, with $m_2\simeq 0.03\,M_\odot$.
The other, ``normal'' binaries have properties more similar to those of the 
disk population of binary millisecond pulsars, with nearly-circular orbits, 
periods $P_b\sim1\,$d (near the short-period end of the distribution for binaries
in the disk) and white-dwarf (WD) companions with $m_2\sin i\simeq 0.2\,M_\odot$.

The large inferred total population of millisecond
pulsars in globular clusters (several hundred in 47~Tuc alone; 
see Camilo et al.\ 2000) and the very high stellar
densities in many cluster cores 
($\rho_c\sim 10^4-10^6\,M_\odot\,{\rm pc}^{-3}$) suggest that
dynamical interactions must play a dominant role in the formation of
these systems. However, the dynamical formation scenarios traditionally
invoked for the production of recycled pulsars in globular clusters
have many problems.

Scenarios based on {\em tidal capture\/} 
of low-mass main-sequence (MS) stars by
neutron stars (NS), followed by accretion
and recycling of the NS
during a stable mass-transfer phase, run into many difficulties.
Serious problems have been pointed out about the tidal capture process
itself (which, because of strong nonlinearities in the regime relevant to
globular clusters, is far more likely to result in a merger than in the
formation
of a detached binary; see, e.g., Kumar \& Goodman 1996; McMillan et al.\ 1990;
Rasio \& Shapiro 1991). Moreover, the basic predictions of 
tidal capture
scenarios are at odds with many observations of binaries and pulsars in
clusters (Bailyn 1995; Johnston et al.\ 1992; Shara et al.\ 1996).
It is likely that
``tidal-capture binaries'' are either never formed, or contribute
negligibly to the production of recycled pulsars.
Physical {\em collisions\/} between NS and red giants have also been
invoked as a way of producing directly NS--WD 
binaries with ultra-short periods (e.g., Verbunt 1987), but
detailed hydrodynamic simulations show that this does not
occur (Rasio \& Shapiro 1991).

The viability of tidal capture and two-body collision
scenarios has become less relevant with
the realization over the last decade that globular clusters contain
dynamically significant populations of {\em primordial binaries\/} (Hut et
al.\ 1992). Dynamical interactions involving
hard primordial binaries are now thought to provide the dominant
energy production mechanism that allows many globular clusters to 
remain in thermal equilibrium and avoid
core collapse over $\sim10^{10}\,$yr (Gao et al.\ 1991; McMillan \& Hut 1994;
Fregeau et al.\ 2003).

In dense clusters, retained NS can acquire binary companions 
through {\em exchange interactions\/}
with these primordial binaries. Because of its large cross section, this
process dominates over any kind of two-body interaction even for low
primordial binary fractions (Heggie, Hut, \& McMillan 1996; Leonard 1989;
Sigurdsson \& Phinney 1993).
In contrast to tidal capture, exchange interactions with
hard primordial binaries (with semimajor axes $a\sim0.1-1\,$AU)
can form naturally the wide
binary millisecond pulsars seen in some low-density globular clusters
(such as PSR B1310$+$18, with $P_b=256\,$d, in M53, which has the lowest
central density, $\rho_c\sim10^3\,M_\odot\,{\rm pc}^{-3}$, of any globular
cluster with observed radio pulsars).
When the newly acquired MS companion, of mass
$\la1\,M_\odot$, evolves up the giant branch, the orbit circularizes
and a period of {\em stable\/} mass transfer
begins, during which the NS is recycled (see, e.g., Rappaport et al.\ 1995).
The resulting NS--WD binaries
have orbital periods in the range $P_b\sim1-10^3\,$d.
However, this scenario
cannot explain the formation of recycled pulsars in binaries with
periods shorter than $\sim1\,$d. To obtain such short periods,
the initial primordial binary must be extremely hard, with $a \la 0.01\,$AU,
but then the recoil velocity of the system following the exchange interaction
would almost certainly exceed the escape speed from the shallow
cluster potential (e.g., $v_e\simeq 60\,{\rm km}\,{\rm s}^{-1}$ for 47~Tuc).

One can get around this problem by considering more carefully
the stability of mass transfer in NS--MS binaries formed through exchange
interactions.
While all MS stars in the cluster {\em today\/} have masses $\la1\,M_\odot$,
the rate of exchange interactions should have peaked at a time
when significantly more massive MS stars were still present.
Indeed, the NS and the most massive primordial binaries will
undergo mass segregation and concentrate in the cluster core on a time scale
comparable to the initial half-mass relaxation time $t_{\rm rh}$. For typical dense
globular clusters, we expect $t_{\rm rh}\sim 10^9\,$yr,
which is comparable to the MS lifetime of a $\sim 2-3\,M_\odot$ star.
If the majority of NS in the cluster core acquire MS companions in the 
range of $\sim1-3\,M_\odot$, 
a very different evolution ensues. Indeed, in this case,
when the MS star evolves and fills its Roche lobe, the
mass transfer for many systems (depending on the mass ratio and evolutionary
state of the donor star) is {\em dynamically unstable\/} and leads to a
common-envelope (CE) phase (see, e.g., Taam \& Sandquist 2000). 
The emerging binary will have a low-mass WD in
a short-period, circular orbit around the NS.

This simple idea has been explored quantitatively using Monte Carlo
simulations of globular cluster dynamics (Rasio, Pfahl, \& Rappaport 2000;
Rappaport et al.\ 2001) and a brief summary of this work is presented in Sec.~2.
A similar scenario, but starting from tidal capture binaries and applied to
X-ray sources in globular clusters, was discussed by Bailyn \& Grindlay (1987).
The possibility of forming
intermediate-mass binaries through exchange interactions was mentioned
by Davies \& Hansen (1998), who pointed out that NS retention in 
globular clusters may
also require that the NS be born in massive binaries. Among
eclipsing pulsars in the disk, at least one system (PSR J2050$-$0827)
is likely to have had an intermediate-mass binary progenitor,
given its very low transverse velocity (Stappers et al.\ 1998).

\section{Results from Dynamical Simulations}

\begin{figure}
\plotone{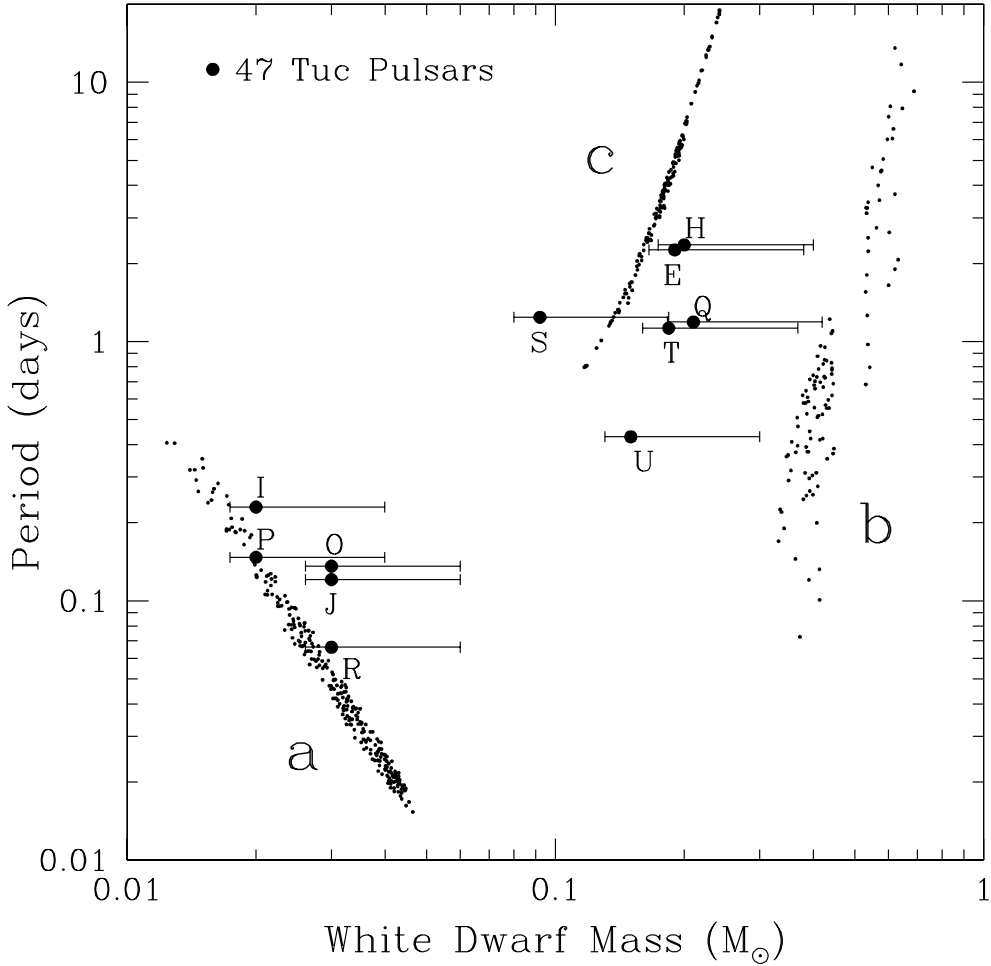}
\caption{Typical results of a dynamical simulation for binary millisecond 
pulsar formation in 47~Tuc.  Each small dot represents a binary
system produced in the simulation, while the filled circles show the
parameters of the observed binary pulsars in 47~Tuc. 
There are 3 separate groups of simulated binaries.  
Systems in the diagonal band on the left (a) come from post-exchange 
binaries that decayed via
gravitational radiation to very short orbital periods ($\sim\,$mins), then
evolved with mass transfer back up to longer periods.  
The sparse group on the right (b) contains post-exchange NS--WD binaries 
that had insufficient time to decay to Roche-lobe contact.  
The NS in this group are not thought to be
recycled since they cannot have accreted much mass during the 
rapid CE phase. Finally, the systems lying
in the thin diagonal band toward longer periods (c) are 
post-exchange binaries in which the mass
transfer from the giant or subgiant to the NS would be stable.  These have not
been evolved through the mass transfer phase; the mass plotted is simply
that of the He core of the donor star when mass transfer commences.  There
are many more systems in this category that have longer periods but lie off
the graph.}
\end{figure}

Figure~1 shows typical results from our most recent numerical simulations
for 47~Tuc (Rappaport et al.\ 2001). Given the many simplifying assumptions
used in this work, the agreement with observations is quite encouraging.
The parameters of observed systems were taken from Camilo et al.\ (2000)
and Freire et al.\ (2001, 2003). We omitted 47~Tuc W, which is now known
to have a MS companion (Edmonds et al.\ 2002). Such systems are the
natural descendants
of longer-period binaries (upper part of group c in Fig.~1), which are 
very likely to interact again in the dense cluster environment.
These multiple interactions can produce both single recycled pulsars 
and binary pulsars with anomalous companions (like 47~Tuc W)
and/or unusual positions in the cluster halo (as in NGC 6752; 
D'Amico et al.\ 2002).

The initial conditions for the simulations
include MS stars and primordial binaries with standard 
initial mass functions and distributions of binary parameters, as well
as a certain number of single NS, which are assumed to have 
been retained by the cluster (cf.\ Pfahl, Rappaport, \& Podsiadlowski 2002). 
The total number of MS stars initially is $\sim 10^6-10^7$ and the total number
of NS is $\sim 10^4$.

Binaries and NS undergo mass segregation and enter
the cluster core in a time 
$t_{\rm s} \sim 10 (m_f/m_t) t_{\rm rh}$, for objects of mass $m_t$
drifting through field stars of average mass $m_f$
(see, e.g., Fregeau et al.\ 2000).  
For the simulation of Fig.~1 we assumed $t_{\rm rh}=10^9\,$yr.
Binaries whose primaries evolve off the MS before entering the 
cluster core are removed from the simulation.
From the numbers of binaries and NS in the core,
we can compute the time for each binary to have a strong interaction
with a NS. We can then decide on a list of actual interactions in each 
timestep using a Monte Carlo procedure. 
All 3--body interactions are computed by direct numerical
integration. Disrupted and ejected binaries are removed from 
the simulation. 

We then calculate the evolution of the newly formed NS--MS binaries.
When the primary evolves off the MS, the orbit is assumed to circularize
(conserving total angular momentum). We then test for the stability
of mass transfer  when the primary fills its Roche lobe.
We find that typically about 50\% of the systems enter a
CE phase. The outcome of the CE phase is calculated using the standard
treatment, with the efficiency parameter $\alpha_{CE}=0.5$
(Rappaport, Di Stefano, \& Smith 1994). A significant fraction of 
these NS--WD binaries will undergo further
evolution driven by gravitational radiation. For orbital periods
$\la 8\,$hr, the companion will be filling its Roche lobe in less
than $\sim10^{10}\,$yr and a second phase of mass transfer will occur.
For WD masses $\la 0.4\,M_\odot$ the mass transfer is stable and
the evolution can be calculated semi-analytically. Our calculations for this
phase also incorporate a simple treatment of the tidal heating of the
companion (Applegate \& Shaham 1994; Rasio et al.\ 2000).
We track the accretion rate and spin-up
of the NS during the mass-transfer phase and we terminate the evolution
when the NS spin period reaches a randomly chosen value in the range 
$2-5\,$ms (at which point the radio
pulsar emission is assumed to turn on and stop the accretion flow).

This simple scenario provides a natural way of explaining the large
number and observed properties of short-period binary pulsars
in a dense globular cluster such as 47~Tuc (Fig.~1).
Although quantitatively the predicted properties of the final binary
population depend on our parametrization of several
uncertain processes (such as CE evolution and tidal heating),
the overall qualitative picture is remarkably robust. Indeed,
quite independent of the details of the various assumptions and
choices of parameters, exchange interactions {\em inevitably\/} form
a large population of NS--MS binaries that will go through a CE phase.
The only way for a globular cluster to avoid forming such a population
would be to start with a very low primordial binary fraction, a
very small number of retained NS, or to have a very long relaxation
time $t_{\rm rh}\ga10^{10}\,$yr, such that all MS stars with masses
$\ga1\,M_\odot$ evolve before the rate of exchange interactions becomes
significant. A large fraction of the post-CE NS--WD binaries cannot
avoid further evolution driven by gravitational wave emission, with
the companion ultimately reduced to a very low mass
$m_2\sim 10^{-2}\,M_\odot$.

Pulsars in wider binaries with $P_b\sim1\,$d and WD
companions must have
evolved from the group of systems with stable mass transfer from a 
$\sim1\,M_\odot$ subgiant 
to a NS, which have orbital periods at the start of mass transfer in
the range $\simeq 1-5\,$d (lower end of group~c in Fig.~1).  
Conventional evolutionary
scenarios suggest that systems where the donor has a
well-developed degenerate core should
inevitably evolve to longer orbital periods.  
However, many of the systems in group~c of Fig.~1 have not yet 
developed such cores, and their final fate is uncertain.  
We also note
that many binary pulsars in the Galactic disk population, which 
are supposed to fit this evolutionary scenario involving stable mass transfer
from a low-mass giant to the NS, have orbital periods shorter 
than $5\,$d, with some $<1\,$d. Even the most sophisticated
binary evolution models currently available cannot explain
these systems (see Podsiadlowski, in this volume).

The results illustrated in Fig.~1 also predict the existence of a large number 
of binary pulsars
with companion masses $m_2\simeq 0.03-0.05\,M_\odot$ and orbital periods
as short as $\sim15\,$min that may have so far escaped detection
(lower end of group~a in Fig.~1).
Future observations using more sophisticated acceleration-search
techniques or shorter integration times may be able to detect
them.

\acknowledgments
I thank Eric Pfahl and Saul Rappaport for many important contributions
to this work.
This work was supported by NASA Grants NAG5-8460, NAG5-11396, and
NAG5-12044.

\end{document}